\documentclass[10pt,letterpaper]{article}
\usepackage{opex3}

\usepackage{amsmath,amssymb,graphicx}
\usepackage{array}
\usepackage{cite}

\newcommand{\de}[2]{\frac{\mathrm{d} #1}{\mathrm{d} #2}}
\newcommand{\eps}{\varepsilon}
\newcolumntype{M}{>{\centering\arraybackslash}m{2.5cm}}
\newcolumntype{S}{>{\centering\arraybackslash}m{1.5cm}}

\begin{document}

\title{Proposal for the generation of photon pairs with nonzero orbital angular 
momentum in a ring fiber}

\author{D. Jav\r{u}rek,$^{1}$ J. Svozil\'{i}k,$^{1,2,*}$ J. Pe\v{r}ina 
Jr.$^{3}$}

\address{$^1$RCPTM, Joint Laboratory of Optics PU and IP AS CR, 17. listopadu 
12, 771 46 Olomouc, Czech Republic\\
$^2$ICFO-Institut de Ci{\`{e}}ncies Fot{\`{o}}niques,
Mediterranean Technology Park, Av. Carl Friedrich Gauss 3, 08860
Castelldefels, Barcelona, Spain\\
$^3$Institute of Physics, Joint Laboratory of Optics PU and IP
AS CR, 17. listopadu 50a, 771 46 Olomouc, Czech Republic}

\email{$^*$jiri.svozilik@upol.cz}

\begin{abstract}
We present a method for the generation of correlated photon pairs
in desired orbital-angular-momentum states using a non-linear
silica ring fiber and spontaneous parametric down-conversion.
Photon-pair emission under quasi-phase-matching conditions with
quantum conversion efficiency $ 6\times 10^{-11} $ is found in a
1-m long fiber with a thermally induced $\chi^{(2)}$ nonlinearity
in a ring-shaped core.
\end{abstract}

\ocis{060.2330,060.5565,190.4975}

\section{Introduction}

Optical fields with non-zero orbital angular momentum (OAM), which
show a non-gaussian transverse amplitude and phase profiles
\cite{Allen2003optical,Torres2011twisted}, are of great interest in a myriad
of scientific and technological applications, such as secure
communications \cite{Bozinovic2013terabit}, ultra-precise
measurements \cite{Molina2007twisted,Puentes2012weak},
nano-particle manipulation \cite{Padgett2011tweezers} or quantum
computing \cite{Zhang2007demonstration}. These applications push
forward the development of new methods aimed at the preparation of
optical fields with OAM in both the classical and quantum (i.e.
single-photon) regimes.

Paired photons with nonzero OAM are widely generated nowadays via
the nonlinear optical process of spontaneous parametric
down-conversion (SPDC), where photons are generated in pairs
(signal and idler). Such paired photons can show quantum
correlations (entanglement) in several degrees of freedom
including polarization, frequency, or momentum. Entanglement may
occur also in the OAM degree of freedom
\cite{Mair2001entanglement}, as experimentally demonstrated in
\cite{Dada2011experimental}. States with the winding numbers
around 300 have already been observed in volume optics
\cite{Fickler2012quantum}. Dimensions of this entanglement
reaching even $ 10^5 $ are theoretically predicted in
\cite{Svozilik2012high}. However, the generation of OAM
entanglement in optical fibers still represents a challenge, which
solution could open the door for many applications that would
benefit from using guided modes (low losses in optical elements,
long transmission distances). Recently, guided OAM states in
fibers with the winding numbers up to 1 have been observed
\cite{Bozinovic2013terabit}.

There are mainly two problems to overcome. On the one hand, the
presence of inverse symmetry in ideally cylindrically-shaped
silicon optical fibers excludes the existence of $\chi^{(2)}$
nonlinearity. For this reason, photon pairs in optical fibers are
generally generated by means of an alternative nonlinear process
(four-wave mixing) which utilizes instead the third-order
nonlinearity of silicon
\cite{Li2005optical,Fulconis2005high,Fan2005efficient}. Small
values of the elements of $\chi^{(3)}$ nonlinear tensor can be
compensated by increasing the interaction length to give higher
photon-pair fluxes. Unfortunately, this is accompanied by equal
enhancement of other effects, i.e., Raman scattering, that cause
unwanted higher noise contributions to the generated flux.
Nevertheless, silicon optical fibers can become nonlinear using
the method of thermal poling
\cite{Myers1991large,Zhu2010measurement} which provides a nonzero
$\chi^{(2)}$ susceptibility and also enables to employ
quasi-phase-matching (QPM) reached via UV erasure
\cite{Bonfrate1999parametric,PhanHuy07photon,Zhu2012direct}.

On the other hand, the propagation of photons with OAM in the
usual step-index long optical fibers do not prevent cross-talk
among modes with different OAM from being strong, which results in
the fast deterioration of the purity of the OAM propagating modes.
However, modern technology suggest also a solution in the form of
ring and vortex fibers with ring-shaped cores \cite{Yue2012mode}
that are more resistant against cross-talk. Here, we show that
entangled photon pairs in OAM modes can be generated in this type
of silica fiber with thermal poling.

\section{Theoretical model}

\begin{figure}[t]   
\centering \hspace{0.05cm}\includegraphics[width=9cm]{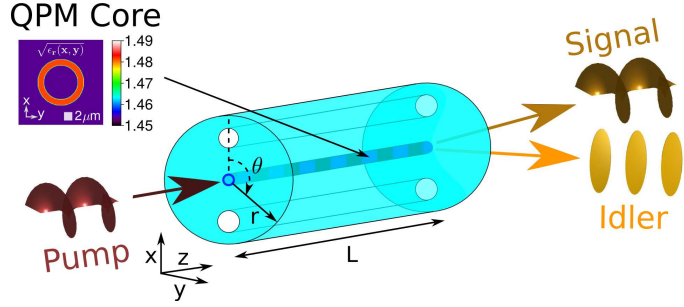}

\caption{Sketch of the proposed silica fiber with a ring-shaped
core utilizing quasi-phase-matching as a source of entangled
photon pairs. The inset represents the transverse profile of
refraction index $ \sqrt{\epsilon_r (x,y)} $ at the pump
wavelength $\lambda_{p0}$ = $0.775\,\mu m$.} \label{Fig:1}
\end{figure}

The profile of the fiber with a ring-shaped core considered here
is shown in Fig.~\ref{Fig:1}. As the optical fields of guided
modes are localized in the close vicinity of the core
\cite{Yue2012mode,Canagasabey2009high}, the resulting radial
symmetry considerably simplifies their determination via the
Maxwell equations. In a real fiber, weak anisotropy occurs due to 
non-homogenity and stress in the material caused by
two holes far away from the ring core that serve for poling wires
\cite{Zhu2010measurement}. Comparison of analytical results with
the precise ones obtained by the full-vector mode solver from
\textit{Lumerical} has revealed that the applied model is
appropriate as long as the elements $ \epsilon_x $ and $
\epsilon_y $ of linear permittivity satisfy the condition
$\sqrt{{\epsilon_y}/{\epsilon_x}}-1 < 1 \times 10^{-7}$.

In rotationally symmetric systems, the full electric (${\bf E}$)
and magnetic (${\bf H}$) fields can be derived from their
longitudinal components $E_z$ and $H_z$. Moreover, they can be
written as the product of a function that depends only on the
azimuthal coordinate ($\theta$), and an another function that
depends only on the radial coordinate ($r$), i.e., $E_z$, $H_z
\sim f(r)g(\theta)\exp[i(\beta z - \omega t)]$. Functions $f(r)$
and $g(\theta)$ describe spatial profiles of fields along the
radial and azimuthal directions, respectively. They satisfy the
eigenvalue equations:
\begin{equation}
 r^2\de{^2 f(r)}{r^2} + r\de{f(r)}{r} + r^2
  \left[\dfrac{\omega^2\epsilon_r}{c^2}-\beta^2 - \frac{n^{2}}{r^2}
  \right] f(r) =0, \hspace{1cm}
 \de{^2 g(\theta)}{\theta^2} + n^2 g(\theta) =0, \label{Eq:1}
\end{equation}
where $\beta$ is the propagation constant (eigenvalue) of a guided
mode with frequency $\omega $; $c$ denotes the velocity of light
in the vacuum and $\epsilon_r$ means the relative dielectric
permittivity. Transverse profile of $\sqrt{\epsilon_r}$ is shown
in the inset of Fig.~\ref{Fig:1}. To simplify our calculations, we
consider $\epsilon_r$ only as a scalar-function. Integer number
$n$ counts eigenmodes that are solutions of the second equation
in~(\ref{Eq:1}) describing azimuthal properties of optical fields.
These solutions have the form of harmonic functions with
frequencies determined by $n$. On the other hand, solutions of the
first equation in~(\ref{Eq:1}) for a fixed value $n$ are expressed
in terms of the Bessel functions. For fixed values of frequency $
\omega $ and index $ n $, the continuity requirements on core
boundaries admit only a discrete set of eigenmodes with suitable
values of propagation constants $\beta $ \cite{Snyder1983optical}.
If we take into account field polarizations, we reveal the usual
guided modes of optical fibers that are appropriate also for
describing photon pairs. Spatial modes characterized by an OAM
(winding) number $l$ are then obtained as suitable linear
combinations of the above fiber eigenmodes $ {\bf
e}_{\eta}(\omega) $ classified by multi-index $\eta$ that includes
azimuthal number $n$, radial index of solutions of the first
equation in~(\ref{Eq:1}) and polarization.

On quantum level, the nonlinear process of SPDC converts a pump
photon into a photon pair described by a first-order perturbation solution of the
Schr\"odinger equation with interaction nonlinear Hamiltonian $ \hat{H}_I $
written in the radial coordinates as follows:
\begin{equation}
 \hat{H}_I(t) = 2\eps_0 \int_{S_{\perp}} rdr d\theta \int_{-L}^{0} dz\,
  \mathbf{\chi}^{(2)}(z): \mathbf{E}_{p}^{(+)}(r,\theta,z,t)
  \hat{\mathbf{E}}_{s}^{(-)}(r,\theta,z,t) \hat{\mathbf{E}}_{i}^{(-)}(r,\theta,z,t)
  + {\rm  h.c.}
\label{Eq:2}
\end{equation}
In Eq.~(\ref{Eq:2}), subscripts $ p $, $ s $ and $ i $ denote in
turn the pump, signal and idler fields. Symbol : is tensor
shorthand with respect to three indices of $\chi^{(2)}$ tensor and
$\epsilon_0$ denotes the vacuum permittivity and $ {\rm h.c.} $
replaces the Hermitian conjugated term. Symbol $S_{\perp}$ means
the transverse area of the fiber of length $L$. Effectively
induced nonlinear susceptibility $ \mathbf{\chi}^{(2)}$ has the
following non-zero elements: $\chi^{(2)}_{xxx} \simeq
3\chi^{(2)}_{xyy}$ and
$\chi^{(2)}_{xyy}=\chi^{(2)}_{yxx}=\chi^{(2)}_{yxy}= 0.021$ pm/V
\cite{Zhu2010measurement}. We note that small variations in the $
\mathbf{\chi}^{(2)}$ profile across the core are effectively
smoothed out by the nonlinear interaction [see Eq.~(\ref{Eq:6})
below]. The fiber is pumped by a strong (classical) pump beam
which positive-frequency part $ {\bf E}_p^{(+)} $ of the
electric-field amplitude can be decomposed into the above
introduced eigenmodes $ {\bf e}_{p,\eta_p}(\omega_p) $ as:
\begin{eqnarray}
 {\bf E}_p^{(+)}(r,\theta,z,t) &=& \sum_{\eta_p} A_{p,\eta_p} \int d\omega_p\,
  {\cal E}_p(\omega_p) {\bf e}_{p,\eta_p}(r,\theta,\omega_p)
  \exp\left(i[\beta_{p,\eta_p}(\omega_p)z-\omega_p
  t]\right).
\label{Eq:3}
\end{eqnarray}
In Eq.~(\ref{Eq:3}), $ A_{p,\eta_p} $ gives the amplitude of mode
$ \eta_p $ and $ {\cal E}_p $ denotes the normalized pump
amplitude spectrum. Similarly, the negative-frequency parts $
\hat{\bf E}_{s}^{(-)} $ and $ \hat{\bf E}_{i}^{(-)} $ of the
signal and idler electric-field operators can be written as:
\begin{equation}
 \hat{\bf E}_{a}^{(-)}(r,\theta,z,t)= \sum_{\eta_a} \int d\omega_a
  \sqrt{\frac{\hbar \omega_a}{4\pi\eps_0 \bar{n}_{a} c}}
  \hat{a}_{a,\eta_{a}}^\dagger(\omega_a){\bf
  e}_{a,\eta_a}^{*}(r,\theta,\omega_a)
    \exp\left(i[\beta_{a,\eta_a}(\omega_a)z-\omega_a
    t]\right),
\label{Eq:4}
\end{equation}
$ a=s,i $ and $ \hbar $ is the reduced Planck constant. Symbol $
\bar{n}_{a} $ stands for an effective refractive index of field $
a $. The creation operators $
\hat{a}_{a,\eta_{a}}^\dagger(\omega_a) $ give a photon into field
$a$ with multi-index $ \eta_a $ and frequency $ \omega_a $.

The state $ |\psi\rangle $ describing a photon pair at the output
plane of the fiber can be written as a quantum superposition
comprising states of all possible eigenmode combinations $
(\eta_s,\eta_i) $:
\begin{eqnarray}
\nonumber
 |\psi\rangle &=& - \frac{i}{c} \sum_{\eta_s,\eta_i} \sum_{\eta_p} A_{p,\eta_p}
  \int d\omega_s \int d\omega_i \, \sqrt{\frac{\omega_s
  \omega_i}{\bar{n}_{s} \bar{n}_{i}}} {\cal E}_p(\omega_s+\omega_i)
  \nonumber \\
 & & \mbox{} \times I_{\eta_p,\eta_s\eta_i}(\omega_s,\omega_i)
  \hat{a}^{\dagger}_{s,\eta_s}(\omega_s) \hat{a}^{\dagger}_{i,\eta_i}(\omega_i)
  |{\rm vac} \rangle;
 \label{Eq:5}
\end{eqnarray}
$ |{\rm vac} \rangle $ is the initial signal and idler vacuum
state. Function $ I_{\eta_p,\eta_s\eta_i}(\omega_s,\omega_i)$
quantifies the strength of interaction among the indicated modes
at the given signal and idler frequencies,
\begin{eqnarray}
 I_{\eta_p,\eta_s\eta_i}(\omega_s,\omega_i) &=&
  \int_{S_{\perp}} rdr d\theta \int_{-L}^0 dz \, \mathbf{\chi}^{(2)}(z) : {\bf
  e}_{p,\eta_p}(r,\theta,\omega_s+\omega_i)
  {\bf e}_{s,\eta_s}^{*}(r,\theta,\omega_s) {\bf e}_{i,\eta_i}^{*}(r,\theta,\omega_i) \nonumber\\
 & & \mbox{} \times
  \exp[i\Delta\beta_{\eta_p,\eta_s\eta_i}(\omega_s,\omega_i)z];
\label{Eq:6}
\end{eqnarray}
$\Delta\beta_{\eta_p,\eta_s\eta_i}(\omega_s,\omega_i) \equiv
\beta_{p,\eta_p}(\omega_s+\omega_i)-\beta_{s,\eta_s}(\omega_s)-\beta_{i,\eta_i}(\omega_i)
$ characterizes phase mismatch of the nonlinear interaction among
individual modes.

Using the state $ |\psi\rangle $ in Eq.~(\ref{Eq:5}), a signal
photon-number density $n_{\eta_s}\left(\omega_s\right)$ observed
in mode $ \eta_s $ is computed along the formula
\begin{eqnarray}
n_{\eta_s}\left(\omega_s\right)&=& \sum_{\eta_i}\int d\omega_i \langle \psi|
\hat{a}^{\dagger}_{s,\eta_s}(\omega_s)\hat{a}^{\dagger}_{i,\eta_i}(\omega_i)
\hat{a}_{s,\eta_s}(\omega_s) \hat{a}_{i,\eta_i}(\omega_i)
|\psi\rangle .
 \label{Eq:7}
\end{eqnarray}

\section{Numerical results}

\begin{figure} 
\centering
(a)\includegraphics[width=6.1cm]{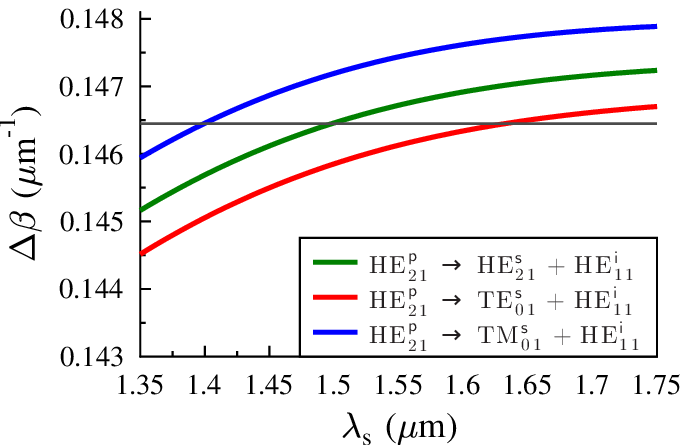}\hspace{0.2cm}(b)\includegraphics[width=6.10cm]{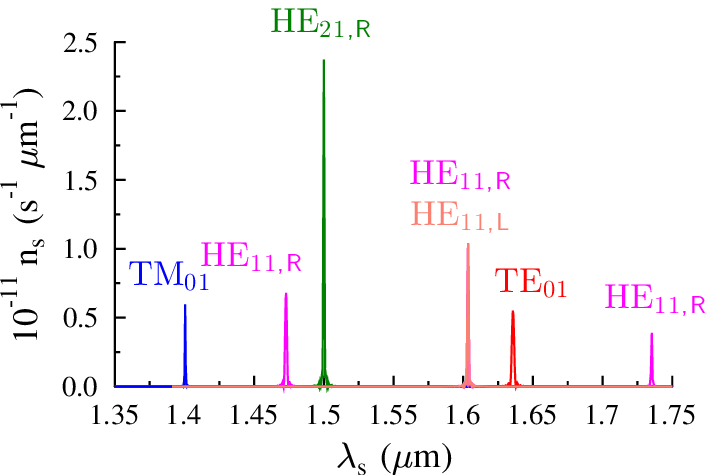}
\caption{
(a) Phase mismatch $\Delta\beta_{\eta_p,\eta_s \eta_i}$ as it
depends on signal wavelength $ \lambda_s $ for three different
combinations of eigenmodes fulfilling the OAM selection rule. The
horizontal grey line indicates the maximum of spatial spectrum of
$ \chi^{(2)} $ modulation expressed in $ - \Delta\beta $. The
period of modulation $ \Lambda = 42.9~\mu $m is chosen such that
quasi-phase-matching occurs for $\lambda_{s0} = 1.5 $~$ \mu $m and
process $\rm{HE}_{21,R}^p\rightarrow {\rm HE}_{21,R}^s,+ {\rm
HE}_{11,R}^i $. (b) Signal photon-number density $ n_{\eta_s}$ as a function
of signal wavelength $ \lambda_s $ for a 1~m-long fiber with
period $ \Lambda = 42.9~\mu $m. Different modes recognized in the
signal field are indicated.} \label{Fig:2}
\end{figure}

In the analysis, we consider a silica fiber with a ring-shaped
core created by doping the base material with 19.3 mol\% of
$\mathrm{GeO_{2}}$ (for details, see \cite{Bruckner2011elements}).
The SPDC process is pumped by a monochromatic beam of wavelength
$\lambda_{p0}=0.775\;\mu$m. The fiber was designed such that
photon pairs are emitted around the wavelength $1.550\;\mu $m used
in fiber communication systems. A right-handed circularly
polarized HE$_{21,R}^p$ mode with OAM number $ l=1 $ (composed of
HE$_{21,\rm odd} $ and HE$_{21,\rm even} $ modes, see
\cite{Snyder1983optical}) has been found suitable for the pump
beam. It gives minimal crosstalk with other pump modes (namely
TM$_{01}$) at the given wavelength. Importantly, according to 
the full-vector numerical model, the propagation constants of 
eigenmodes HE$_{21,\rm odd} $ and HE$_{21,\rm even} $ for pump (signal) field 
differ only by 3.88$\times10^{-3}$ (1.14$\times10^{-3}$ ) rad/m in 
an anisotropic fiber with $ \sqrt{{\epsilon_y}/{\epsilon_x}}-1 = 1 \times 
10^{-7}$. The
generated signal and idler photons fulfil the energy conservation
law ($\omega_{p}=\omega_{s}+\omega_{i}$) and also the selection
rule for OAM numbers ($l_{\eta_p}=l_{\eta_s}+l_{\eta_i}$) that
originates in the radial symmetry of the nonlinear interaction.
Under these conditions, efficient photon-pair generation has been
found for a signal photon in mode HE$_{21,R}^s$ ($l_s=1$) and an
idler photon in mode HE$_{11,R}^i$ or HE$_{11,L}^i$ ($l_i=0$)
which represent the right- and left-handed polarization variants
of the same spatial mode. As the phase-matching curves in
Fig.~\ref{Fig:2}(a) show, also other efficient combinations of
signal and idler modes are possible, namely HE$_{21,R}^p
\rightarrow $ TE$_{01}^s$ $+$HE$_{11}^i$ and HE$_{21,R}^p
\rightarrow $ TM$_{01}^s$ $+$ HE$_{11}^i$.

However, QPM reached via the periodic modulation of $\chi^{(2)}$
nonlinearity allows to separate different processes. The right
choice of period $\Lambda$ of $\chi^{(2)}$ nonlinearity tunes the
desired process that is exclusively selected provided that the $
\chi^{(2)} $ spatial spectrum is sufficiently narrow. For our
fiber, the $ \chi^{(2)} $ spatial spectrum has to be narrower than
$ 1 \times 10^{-3}$ $\mu $m${}^{-1}$. This is achieved in general
for fibers longer than 1~cm. The analyzed fiber 1~m long with the
width of spatial spectrum equal to $ 7.6 \times 10^{-6}$ $\mu $m$
{}^{-1}$ allows to separate the desired process from the other
ones with the precision better than 1:100.

According to Fig.~\ref{Fig:2}(b) the greatest values of signal
photon-number density $ n_{s,\eta_s} $ occur for mode
HE$_{21,R}^s$ with an OAM number $l=+1$ around the wavelength
$\lambda_s=1.5~\mu $m. The full width at the half of maximum of
the peak equals $\Delta \lambda_s = 0.96$~nm. The second largest
contribution belongs to the processes involving modes
HE$_{11,R}^s$ and HE$_{11,L}^s$ that interact with mode
HE$_{21,R}$. They build a common peak found at the wavelength
$\lambda_s=1.603 $~$\mu $m. In signal photon-number density $
n_{s,\eta_s} $, there also exists two peaks of mode HE$_{11,R}^s$
that form pairs with the peaks created by modes TM$_{01}$ and
TE$_{01}$. These peaks are shifted towards lower and larger
wavelengths, respectively, due to their propagation constants.
Whereas the peak belonging to mode TM$_{01}^s$ occurs at the lower
wavelength $\lambda=1.4$~$ \mu $m, the peak given by mode
TE$_{01}^s$ is located at the longer wavelength $\lambda=1.635$~$
\mu$m. Spectral shifts of these peaks allow their efficient
separation by frequency filtering. The generated photon-pair field
is then left in the state with a signal photon in mode
HE$_{21,R}^s$ and an idler photon either in state HE$_{11,R}^i$ or
HE$_{11,L}^i $. The weights of both possible idler states
HE$_{11,R}^i$ and HE$_{11,L}^i $ in quantum superposition are the
same which gives linear polarization of the overall idler field.
Also, pump-field leakage into modes TE$_{01}^p$ and TM$_{01}^p$
caused by imperfect fiber coupling may occur. This gives extra
peaks in the signal photon-number density $n_{s,\eta_r}$ that are
spectrally separated from those discussed above by 15~nm.

The fiber 1~m long provides around 240 photon pairs per 1~s and $
1~\mu $W of pumping in the analyzed modes. For comparison, the
fibre 10-cm long generates around 20 photon pairs per 1~s and $
1~\mu $W of pumping in the same configuration. This means that a
slightly better than linear increase of photon-pair numbers with
the fiber length occurs. Strong correlations between the signal
and idler frequencies result in fast temporal correlations between
the signal and idler detection times occurring in a time window
7~ps long. We notice, that the process can also be considered in its
left-handed polarization variant in which the pump beam propagates
as a HE$_{21,L} $ mode. Both variants provide photon pairs
suitable, e.g., for quantum metrology or heralded single-photon
sources \cite{Collings2013Integrated} giving Fock states with
non-zero OAM numbers.

Quasi-phase matching allows also other efficient combinations of
modes. For example, the pump beam in mode HE$ {}_{11,R}^p $ (or
HE$ {}_{11,L}^p $) with $ l_p = 0 $ provides spectrally broad-band
SPDC that may give photon pairs with temporal correlations at fs
time-scale. Also photon pairs entangled in OAM numbers can be
obtained in this configuration in which the signal (idler) photon
propagates either as HE$^{s}_{21R}$ (HE$^{i}_{21L}$) mode or
HE$^{s}_{21L}$ (HE$^{i}_{21R}$). We note that vortex fibers
\cite{Bozinovic2013terabit} are also suitable for SPDC as they
provide similar conditions for photon-pair generation as the
analyzed ring fibers. Moreover, their additional core gives better
stability to fundamental modes participating in the nonlinear
interaction. Both ring and vortex fibers thus have a large
potential to serve as versatile fiber sources of photon pairs in
OAM states useful in optical information processing.

\section{Conclusion}
A ring fiber with thermally induced $ \chi^{(2)}$ nonlinearity and
periodical poling has been presented as a promising source of
photon pairs being in eigenmodes of orbital angular momentum.
Spontaneous parametric down-conversion has been pumped by a beam
with nonzero orbital angular momentum that has been transferred
into one of the down-converted beams. Several mutually competing
nonlinear processes exploiting different modes can be spectrally
separated. Other configurations also allow for the emission of
spectrally broad-band photon pairs as well as photon pairs
entangled in orbital-angular-momentum numbers. This makes the
analyzed ring fiber useful for many integrated fiber-based
applications.

\section*{Acknowledgments}

The authors thank J.P. Torres for his advice and useful comments.
D.J. acknowledges J.P. Torres for his support at ICFO. Support by
projects CZ.1.05/2.1.00/03.0058 and CZ.1.07/2.3.00/20.0017 of
M\v{S}MT \v{C}R and P205/12/0382 of GA \v{C}R are acknowledged.
D.J. and J.P. thanks the project PrF\_2013\_006 of IGA UP Olomouc.
J.S. thanks the projects CZ.1.07/2.3.00/30.0004 and
CZ.1.07/2.3.00/20.0058 of M\v{S}MT \v{C}R.

\end{document}